\documentclass[12pt,amsmath,amssymb]{iopart}
\usepackage{epsf}

\newcommand{\ve}[1]{\bi{#1}}

\begin{document}

\title[Electron-phonon coupling in a two-dimensional inhomogeneous electron gas]{Electron-phonon coupling in a 
two-dimensional inhomogeneous electron gas: consequences for 
surface spectral properties}
\author{Natalia Pavlenko$^{1,2}$ and Thilo Kopp$^1$}

\address{
$^1$ Center for Electronic Correlations and Magnetism, Universit\"at Augsburg, 86135 Augsburg, Germany\\
$^2$ Institute for Consensed Matter Physics, 79011 Lviv, Ukraine}

%{Inhomogeneous superconductors and superconducting systems}
%\pacs{74.78.-w}{Superconducting films and low-dimensional structures}
%\pacs{73.20.Mf}{Collective excitations}

%\newcommand{\ve}[1]{\boldsymbol{#1}}

\begin{abstract}
We investigate the coupling of an inhomogeneous electron system to phonons.
The properties of an electronic system composed of a mixture
of microscopic ordered and disordered islands are changed fundamentally by a phonon mode.
In high-$\boldmath{T_c}$ cuprates, such
a phase separation scenario is supported by recent STM
and nuclear quadrupole resonance studies. We show that even a weak or moderate electron-phonon
coupling can be sufficient to produce dramatic changes in the electronic state of the 
inhomogeneous electron gas. 
The spectral properties calculated in our approach provide a natural explanation of
the low-energy nodal ARPES
features and exhibit a novel non-Fermi-liquid state stabilized through
electron-phonon coupling. 
\end{abstract}

\submitto{\JPCM}
\maketitle

\section{Introduction}

An eminent feature observed in the angle-resolved photoemission spectroscopy
(ARPES) studies of high-$\boldmath{T_c}$ cuprates
is the kink-like change of electron velocity,
interpreted in terms of the coupling
to oxygen phonon modes \cite{valla,lanzara}.
These kinks are associated with
a distinct break-up of the spectral weight into a high-intensity part which develops
near the Fermi surface, and a broad
structure of less intensity at higher energies
below the Fermi level.
Despite the structural differences between various types of cuprates,
the peculiarities in the dispersion at 50--80~meV
were detected in Bi$_2$Sr$_2$CaCu$_2$O$_{8+\delta}$ (Bi2212),
La$_{2-x}$Sr$_x$CuO$_4$ (LSCO), YBa$_2$Cu$_3$O$_{6+x}$ (YBCO) and other related
systems \cite{valla,lanzara,zhou,kordyuk,borisenko,damascelli}, and are
believed to shed light on the microscopic mechanism of high-$T_c$
superconductivity. Although significant efforts have been directed towards a theoretical
investigation of the influence of many-body interactions
on the one-electron properties of Hubbard and $t$--$J$ models, typically used
for cuprates \cite{yunoki}, the origin of the nodal ARPES features
and of their unusual doping and temperature behavior
still remains an open question.

Many-body effects are known to produce changes in the electronic dispersion, as
deduced from electronic photoemission spectra. One mechanism, which was shown
to cause low-energy kinks, is controlled by a strong electron-phonon coupling
in the free electron gas \cite{hengsberger,valla2}. This coupling leads to a renormalization of the
electron effective mass
in the energy range close to the Fermi energy $E_F$, below a
characteristic phonon frequency $\omega_{\rm ph}$: $E-E_F<\omega_{\rm ph}$. On the other hand, when the
electrons in a metal are strongly correlated, a purely electronic mechanism can
also lead to dispersion kinks which have been related to a crossover between
Fermi- and non-Fermi-liquid behavior \cite{byczuk}.

In the analysis of ARPES intensities, one should always consider the fact that
photoemission experiments analyze electronic surface and subsurface states
which brings the near-surface correlations into the focus.
Due to the strong interaction of these electrons with subsurface
phonon modes and virtual charge transfer excitations,
the effect of such collective modes
on the electronic subsystem is crucial \cite{pavlenko,brown}. A direct consequence of
these interactions is a significant reduction of the local Hubbard repulsion to
values well below the electronic bandwidth $8t$, where $t$ is the electron
hopping. This finding allows to suggest that mechanisms different from the purely
electronic could play a role in the appearance of the dispersion kinks in cuprates.
As the coupling with magnons cannot satisfactorily explain the doping and temperature
behavior of such kinks, we will focus on the analysis of the interaction with phonons.

In the high-$T_c$ cuprates another important factor, which has to be addressed
in studies of their electronic properties, is
charge inhomogeneity. In the pseudogap state and in regimes with suppressed superconductivity,
scanning tunneling microscopy (STM) experiments indicate a local electron order
\cite{kapitulnik,kivelson,vershinin,kohsaka,tranquada,hoffman,hanaguri,lee}. This,
together with nuclear quadrupole resonance (NQR) and
resonant soft X-ray scattering studies \cite{pan,singer,ofer,abbamonte},
provides strong support for a state of electronic phase separation as one of
the widely discussed scenarios for under- and optimal-doped cuprates
\cite{kapitulnik,vershinin,kohsaka,pan,singer,ofer}. Such a state exhibits a
mixture of microscopic charge ordered (characterized as ``more insulating''
with suppressed local density of states) and charge disordered uniform (metallic)
domains with a dominance of the metallic phase at higher doping levels
\cite{kohsaka}.
The ARPES intensities obtained for the electronic
phase separated state inevitably
contain the contributions from both types of domains. This fact can be easily
understood since the electrons collected on the ARPES analyzer can be emitted
from the disordered as well as from the ordered surface islands. Due to the
quantum nature of the collected electrons, they cannot be described in terms of
pure ``ordered'' or ``disordered'' electron wave functions, but rather as a
superposition of both states. Therefore, in order to understand ARPES
intensities within such an inhomogeneous scenario, we need to analyze the
consequences of the electron-phonon coupling not only for the disordered, but
also for the ordered sections of the surface.

In the hole-doped cuprates, the recent ab-initio studies have demonstrated a weakness
of the electron-phonon coupling which should result in a negligibly small 
contribution to the formation of the dispersion kinks \cite{giustino,manske}. Despite the different approaches,
these works have been focused on the electronic homogeneous state, without any 
serious attention to a possible electronic charge order. As a consequence of
such a predomimant consideration of the uniform electronic state, the general view about a relative
unimportance of the electron-phonon coupling for the electronic properties of
the copper oxide planes became widely accepted in the literature. In an alterating 
approach, the strongly inhomogeneous interactions of the electrons to several optical 
phonon modes (buckling and breathing) has been proposed \cite{cuk} which 
was not really successful in the explanation of real physical mechanisms of spectral 
anomalies in the cuprates. 

In the present work, we analyze the electron-phonon coupling in the charge-ordered state.
We show that in contrast to the uniform electron gas, in the ordered system even a weak
or moderate coupling to an optical phonon mode produces dramatic changes in the electronic
properties and leads to a formation of new electronic state which cannot be described
by a standard Fermi-liquid theory. Due to significant advances achieved 
in the recent STM studies of the cuprates \cite{kapitulnik,kohsaka2}, the existence of
local electron order in these systems becomes rather a well established fact. 
To make a step towards an interpretation of the spectral features in the context of
the inhomogeneous state observed in the STM studies, the inhomogeneous surface
in our work is described as a mixture of the electronically uniform and ordered states.
We obtain that the superposition of these states produces several distinct features 
and a characteristic intensity break-up in the electronic spectral maps. In our work,
we connect this break-up with the spectral anomalies observed in the cuprates. Consequently,
the proposed approach is expected to shed new light on the mechanisms of kinks and on the
origin of non Fermi-liquid behavior detected in these systems.

\section{Electronic spectral properties in the charge-ordered state}

To gain deeper insight into the ordered surface state, we provide a comparison
of a disordered two-dimensional electron gas, characterized by a free tight-binding dispersion
$\varepsilon_0(\ve{k})$, with an ordered
electron system, both coupled to an optical phonon mode with a
frequency $\omega_{\rm ph}=\omega_0$. In our studies, we take into account the Coulomb interaction
$V$ between nearest neighbors
\begin{eqnarray}
H_{\rm el}=\Sigma_{\ve{k},\sigma} \varepsilon_0(\ve{k}) n_{\ve{k}\sigma}+V\Sigma_{\langle ij \rangle}
n_i n_j
\end{eqnarray}
where $n_{\ve{k}\sigma}$ are the electron number operators and $n_i=\Sigma_{\sigma} n_{i\sigma}$.
The electron-phonon interaction is considered in terms of a Holstein approach
\begin{eqnarray}
H_{\rm el-ph}=-g\Sigma_i n_i(b_i+b_i^{\dag})+\omega_0\Sigma_i b_i^{\dag}b_i ,
\end{eqnarray}
where the phonon operators refer to a vibration mode of frequency $\omega_0$.
The parameter $g=\sqrt{\omega_0 E_p}$ ($E_p$ is the polaron binding energy) refers the coupling
of the holes in the copper oxide planes to the motion of apical oxygens
in the top surface planes of the samples. In our calculations, the low-energy 
phonon frequency $\omega_0=0.05-0.1t$ reflects the softening of the surface optical phonon
modes suggested in Ref.~\cite{brown}. 
For the free electronic dispersion $\varepsilon_0(\ve{k})=-2t\eta_{\ve{k}}^+-\xi_{\ve{k}}$
(where $\eta_{\ve{k}}^+=\cos k_x+\cos k_y$, $\xi_{\ve{k}}=\mu+4t'\cos k_x \cos k_y$ and
$\mu$ is the chemical potential) we
choose $t=0.18$~eV, $t'=-0.4t$, which is in the range of typical values found
from fitting ARPES data for Bi2212 and Tl2201 \cite{eschrig,plate}.

\begin{figure}[ht]
\epsfxsize=6.5cm \centerline{\epsffile{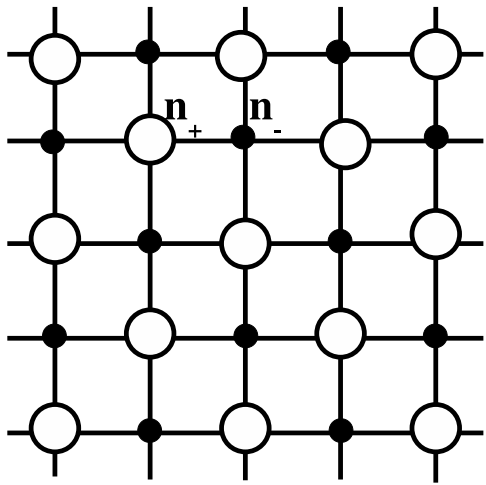}} \caption{Schematic view of checkerboard ording in
a two-dimensional electron gas.} \label{fig1}
\end{figure}

\begin{figure}[ht]
\epsfxsize=8.2cm \centerline{\epsffile{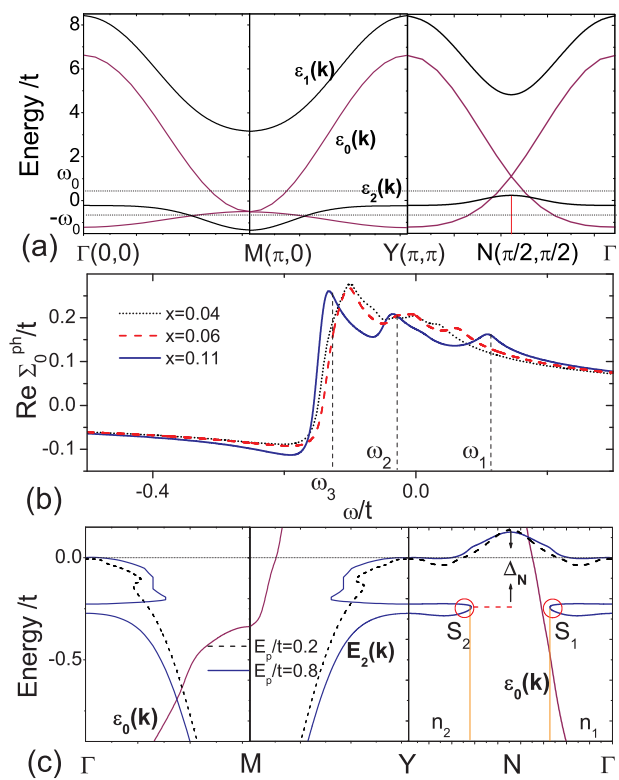}} \caption{Electronic band structure and
self energy for a charge ordered system coupled to phonons where $V/t=1.3$. 
(a) Band structure
with vanishing phonon coupling in the ordered and charge disordered uniform
states, where the dotted lines refer to $\pm \omega_0=0.5t$. Here $kT/t=0.03$ and
$x=0.11$. (b) Real part of $\Sigma_0^{ph}(\ve{k},\omega)$ calculated in the
ordered state with $\omega_0=0.05t$ for different doping levels. Here $\ve{k}$ is
located in the nodal region close to the point ${\rm N}=(\pi/2, \pi/2)$ of the
Brillouin zone; $|\ve{k}|=0.9$, $kT/t=0.03$ and $E_p=1.2t$. (c) Evolution of
the renormalized low-energy band structure in the ordered state with increasing
$E_p$ for $x=0.11$ and $\omega_0=0.05t$. The transition at $E_p=E_p^*\approx 0.2t$ is characterized
by a transformation of the local maximum $E_2(\ve{k}={\rm N})$ into a
hat-shaped structure and by the buildup of a singular $E_2(\ve{k})$ in the
regions $n_1$ and $n_2$ with singular electron velocity $v$. The regions
$\ve{k}\sim \ve{k}_F$ are separated by a gap $\Delta_N$ from $\ve{k}\sim {\rm
N}$ with hole-like excitations. The Fermi level is indicated by dots.}
\label{fig2}
\end{figure}

\begin{figure}[ht]
\epsfxsize=8.5cm \centerline{\epsffile{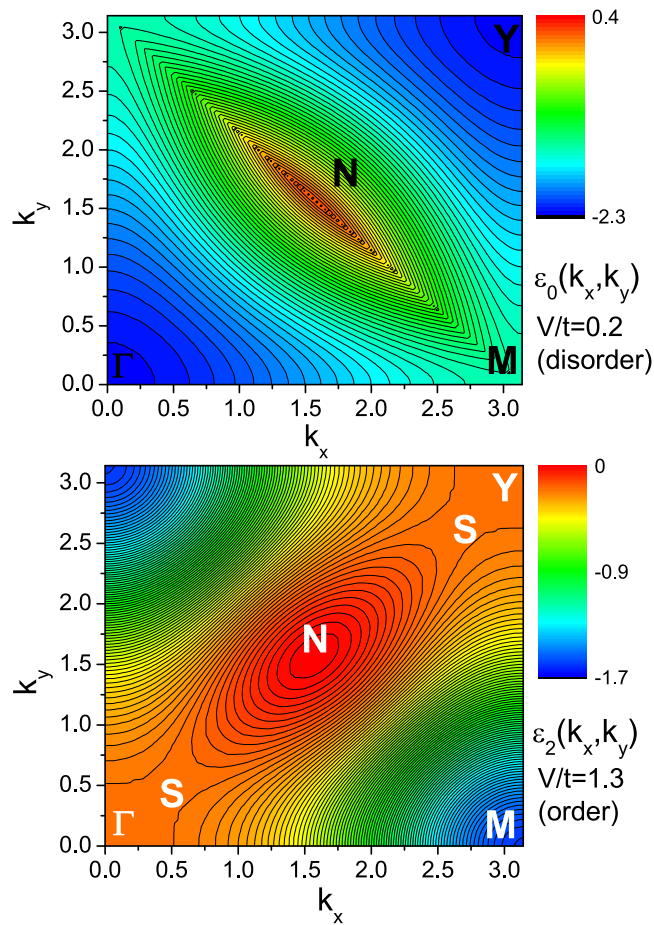}} \caption{Electronic contour plots of
$\varepsilon_0(\ve{k}_x,\ve{k}_y)$ and $\varepsilon_2(\ve{k}_x,\ve{k}_y)$
in the disordered ($V/t=0.2$) and ordered ($V/t=1.3$) state where 
the local extremal points are indicated by $\Gamma$, M, Y, N, and S.  
Here $kT/t=0.03$, $x=0.11$, $E_p/t=1.2$.} \label{fig3}
\end{figure}

In our analysis we consider a checkerboard electronic ordering
which introduces a doubling of the square unit cell shown in Fig.~\ref{fig1}.
This ordering is parameterized in terms of sublattice electronic occupancies $n_{\pm}$ with
order parameter $\delta=n_+-n_-$ and average electron concentration $(n_+
+ n_-)/2=1-x$. The values of $\delta$ are obtained from the minimization of the mean-field type
free energy.
In such a charge ordered state, the electronic band structure is characterized
by two subbands $\varepsilon_j(\ve{k})=-\xi_{\ve{k}}+4Vn\pm \Delta_{\Sigma}$.
The correction $\Delta_{\Sigma}=\sqrt{(2V\delta)^2+(2t\eta^+_{\ve{k}}-\Sigma_{+-}^{HF})^2}$
originates from intersublattice electron correlations,
introduced through the Hatree-Fock self-energy
$\Sigma_{+-}^{HF}=-\int d\ve{q} \sum_{\omega_{n}}
V\cdot \eta_{\ve{k}-\ve{q}}^+\cdot G_{+-}^{HF}(\ve{q},\omega_{n})$
in the Matsubara electron propagators $G^{HF}_{\alpha\beta}(\ve{q},\omega_{n})$.
The new feature arising in the ordered band structure is the flatness of
the energy subbands $\varepsilon_j(\ve{k})$ which is generated by their
splitting through the charge-order gap $\Delta_0=4V\delta$, Fig.~\ref{fig2}(a).
Here, in contrast to the disordered free dispersion
$\varepsilon_0(\ve{k})$, the emergence
of the gap $\Delta_0$ results in the formation of new local extrema of the
ordered dispersions $\varepsilon_j(\ve{k})$. Fig.~\ref{fig3} shows a comparison of the
detailed ($k_x$, $k_y$) map of the low-energy ordered subband $\varepsilon_2(\ve{k})$ 
with the corresponding map of the free electron dispersion $\varepsilon_0(\ve{k})$.
One can immediately see that the electron order produces dramatic changes in the topology 
of the electronic structure which include (i) the formation of a new maximum in the
central nodal point ${\rm N}=(\pi/2,\pi/2)$ of the Brillouin zone; (ii) the appearance
of new saddle points ${\rm S}=(k_x^S, k_y^S)$.   
The saddle point S is located close to $\Gamma$ (the second symmetric saddle point
is close to Y) in the nodal direction so that
$k_x^S=k_y^S=k_S$ where the coordinate $k^S$ is given by
\begin{eqnarray}\label{ks}
\cos k_S=\frac{\sqrt{(4t+2VI_{+-})^4-(16t'V\delta)^2}}{8|t'|(4t+2VI_{+-})},
\end{eqnarray}
where the quantity $I_{+-}$ parametrizes the Hartree-Fock self-energy 
$\Sigma_{+-}^{HF}(\ve{k})=-(V/2)\eta_{\ve{k}}^+I_{+-}$. For the chosen values of the model parameters,
the maximal values of $I_{+-}$ are of the order $0.5$. The expression (\ref{ks})
clearly shows that the saddle point ${\rm S}$ is controlled by the charge order gap and disappears
for weaker Coulomb repulsion $V$,
%% and --- hier meinte ich, dass es etwas expliziter ausgedrueckt werden sollte: "i.e., it is absent ..."
i.~e., it is absent in the disordered uniform state 
with $\delta \rightarrow 0$. We also note that the nodal point ${\rm N}$ indicated in the top
map of the band $\varepsilon_0(\ve{k})$ in Fig.~\ref{fig3} corresponds in fact to the intersection
point of the two branches of $\varepsilon_0(\ve{k})$ (see Fig.~\ref{fig2}(a)).
%% das wird zu lang, deshalb der Punkt und die vorgeschobene Referenz zur Figur.
They are generate with the convolution to the first Brillouin zone and 
%% does not represent a real maximum (see Fig.~\ref{fig2}(a)).  
do not show an extremum at ${\rm N}$. In contrast 
to $\varepsilon_0(\ve{k})$, the new local extrema ${\rm N}$ and ${\rm S}$ 
of $\varepsilon_2(\ve{k})$ are well-defined and should be considered as {\it a
generic feature of the band structure in the electron-ordered state}.

When the flat low-energy subband $\varepsilon_2(\ve{k})$ is coupled to
the phonon mode, the local maximum at ${\rm N}$ and the saddle point at ${\rm S}$ 
lead to van-Hove
singularities in the sublattice contributions to the electronic
self-energies $\Sigma_{\alpha}^{ph}=-g^2\int d\ve{q} \sum_{\omega_{n'}}
G_{\alpha \alpha}^{HF}(\ve{k}-\ve{q},\omega_n-\omega_{n'}) D(\ve{q},\omega_{n'})$.
Here $D(\ve{q},\omega_{n'})=
{2\omega_0}/{((i\omega_{n'})^2-\omega_0^2)}$ is the unperturbed phonon Green function\footnote{
As the main results for the electronic properties, including
the topological anomalies of the dispersion and the NFL-state, appear already at small
electron-phonon coupling $E_p/t \sim 0.2$,
undressed phonon Green functions have been used in the calculations of
the electronic self-energy.}
%\cite{phonon_g},
and the one-electron sublattice propagators $G^{HF}_{\alpha \alpha}$ ($\alpha=\{+,-\}$)
are calculated in the ordered electron state in the self-consistent Hartree-Fock approximation.
Furthermore, the phonon scattering term can be
decomposed as $\Sigma_{\pm}^{ph}=\Sigma_0^{ph}\mp \Delta \Sigma_0^{ph}$ where the
part $\Delta \Sigma_0^{ph} \sim V \delta$ disappears in the charge disordered state.
With this form of $\Sigma_{\pm}^{ph}$, the renormalized electronic propagators can be 
conveniently presented
as a combination of the band Green functions $g_j=1/(i\omega_n+\xi_{\ve{k}}-\Sigma_j(\ve{k},\omega_n))$:
$G_{\alpha \alpha}=G_0 \mp \Delta G_0$, where $G_0=(g_1+g_2)/2$ and
$\Delta G_0=(2V\delta+\Delta \Sigma_0^{ph})(g_1-g_2)/(\Sigma_1-\Sigma_2)$.
In the band propagators $g_j$, the effective self-energy parts
$\Sigma_j=
\Sigma_0^{ph}+4Vn \pm \sqrt{(\Delta{\Sigma}_{0}^{ph}+2V\delta)^2+(2t\eta^+_{\ve{k}}-\Sigma_{+-}^{HF})^2}$
contain the information about both, charge density fluctuations and electron-phonon scattering processes.
We note that $\Sigma_2$ and $g_2$, which determine the low-energy
quasi-particle excitations, are of prime importance.

\begin{figure}[ht]
\epsfxsize=8.5cm \centerline{\epsffile{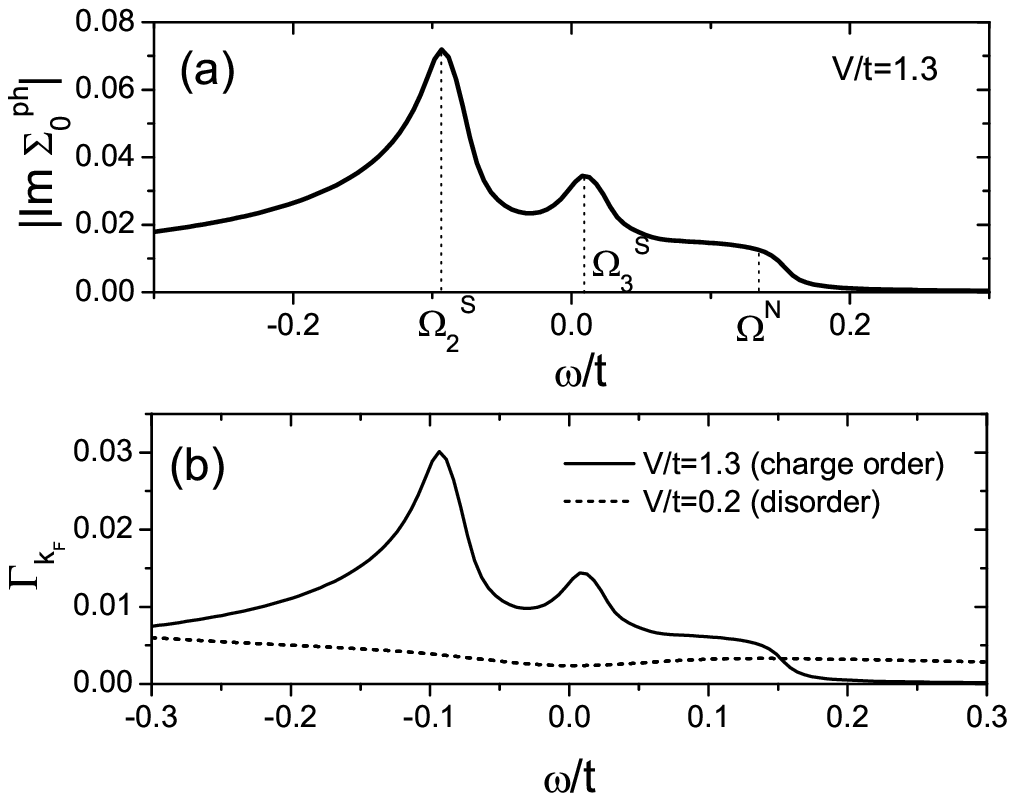}} \caption{(a) Imaginary part of
$\Sigma_0^{ph}(\ve{k},\omega)$ and (b) scattering rate
$\Gamma_{\ve{k}_F}=
Z_{\ve{k}_F}|{\rm Im}\Sigma_0^{ph}|$ ($Z_{\ve{k}_F}$ is the quasiparticle residue)
in the charge ordered state (case $V/t=1.3$,
$|\ve{k}|=1.11$)
and for the free electron gas ($V/t=0.2$, $|\ve{k}|=1.3$).
Here $\ve{k}$ is located in the nodal direction near the ${\rm
N}$-point of the Brillouin zone, $kT/t=0.03$, $x=0.11$, and
$E_p/t=0.2$.}
\label{fig4}
\end{figure}

In Fig.~\ref{fig2}(b), we show the frequency-dependent real part of
$\Sigma_0^{ph}(\ve{k},\omega)$ at different doping levels $x$. The kinks appearing in
$\Sigma_0^{ph}$ at $\omega=\omega_i$ ($i=1,\ldots,3$) correspond to the van-Hove singularities
in $\Sigma_{\pm}^{ph}$. In the charge disordered state, the typical van-Hove singularities
in $\Sigma_0^{ph}$ are caused by the local extrema in the electron dispersion $\varepsilon_0(\ve{k})$
which appear in the high-symmetry points $\Gamma$, ${\rm Y}$ and ${\rm M}$ of the
Brilloiun zone (Fig.~\ref{fig2}(a)). In the charge ordered state, the flattening of $\varepsilon_2(\ve{k})$
produces the additional nodal maximum of
$\varepsilon_2(\ve{k}={\rm N})$ and the saddle point at $\ve{k}={\rm S}$
which lead to the appearance
of new van-Hove singularities in $\Sigma_{\pm}^{ph}$ at small $\omega$.
These  low-energy singularities correspond to the peaks in ${\rm Im} \Sigma_{0}^{ph}$
(see Fig.~\ref{fig4}) and are related to the kinks in ${\rm Re} \Sigma_{0}^{ph}$ (Fig.~\ref{fig2}(b)).
In the ordered state, the energies
of the low-frequency singularities are determined by the equations
$\omega-\varepsilon_2(\ve{k}_i)\pm \omega_0=0$ where $\ve{k}_i=\{ \Gamma, {\rm N}, {\rm S}\}$.
Specifically, while the maximum at $\ve{k}={\rm N}$ produces a jump of $\Sigma_{0}^{ph}$
described by ${\rm Im}\Sigma_{0}^{ph} \sim \Theta(\Omega^{\rm N}-\omega)$ at small
$\Omega^{\rm N}=\varepsilon_2(\ve{k}={\rm N})+\omega_0> 0$,
the saddle point ${\rm S}$ leads to a distinct logarithmic singular behavior of $\Sigma_{0}^{ph}$:
\begin{eqnarray}
{\rm Im}\Sigma_{0}^{ph}\sim g^2 [(1+b_0-f(\omega-\omega_0))\log|(\Omega_2^S-\omega)/t| \nonumber \\
+(b_0+f(\omega+\omega_0))\log|(\Omega_3^S-\omega)/t|],
\end{eqnarray}
where $b_0=b(\omega_0)$ and $f(\omega)$ are the Bose and Fermi distributions functions, and
the energies $\Omega_{2/3}^S=\varepsilon_2(k_S)\mp \omega_0$ are located close to
the Fermi level: $\omega_3< \Omega_2^S< \omega_2$,
$\omega_2< \Omega_3^S < 0$.

As the chemical potential directly affects $\varepsilon_2$ in these equations,
the kinks of $\Sigma_2(\ve{k},\omega)$ at $\omega=\omega_i$ which result from the new
van-Hove singularities are strongly
doping-dependent. It is noteworthy that these low-frequency kinks at $\omega=\omega_i$
($i=1,\ldots,3$) shown in Fig.~\ref{fig2}(b)
will inevitably change the one-electron spectral properties near the Fermi level. To
see this, we present in Fig.~\ref{fig2}(c) the low-energy dispersion
$E_2(\ve{k})$ of the underdoped system ($x=0.11$), now calculated
from the equation $\omega+\xi_{\ve{k}}-\Sigma_2(\ve{k},\omega)=0$ for the poles of the
renormalized electron propagators
for different values of electron-phonon coupling. A central
property resulting from the new van-Hove singularities is a topological
reconstruction of $E_2(\ve{k})$ close to the ${\rm N}$-point. In
Fig.~\ref{fig2}(c), (case $E_p=0.8t$), this reconstruction corresponds to a
transformation of the maximum $E_2(\ve{k}={\rm N})$ into a new singular
hat-shaped band region which appears above a critical value $E_p^*\approx 0.2t$.

The novel nodal dispersion exhibits a nonanalytic character in the symmetric
regions $n_1$ and $n_2$ near the Fermi-level. In these regions, $E_2(\ve{k})$
is a multi-valued function of $\ve{k}$. The two extra subbranches of $E_2$ emerge from the
additional poles of the one-electron Green function and are caused by the
van-Hove kinks in $\Sigma_0^{ph}(\ve{k},\omega)$. In the singular points $S_1$ and $S_2$ of
the regions $n_1$ and $n_2$ which
connect monotonically decreasing and increasing branches of $E_2(\ve{k})$, the nodal electron velocity
$v=|\nabla_{\ve{k}}E_2(\ve{k})|$ approaches infinite values.
The range $\ve{k}$ close to $\ve{k}_F$ with singular $v$ is separated from
the region $\ve{k}\approx {\rm N}$ (characterized by hole-like excitations)
through a gap $\Delta_N/t\approx 0.2$. These topological anomalies are especially
significant in the under- and optimal-doped range and for polaron energies
above $E_p^*$. The nodal properties of the new electronic state,
stabilized at $E_p=E_p^*$, cannot be classified in terms of Fermi liquid
theory. 

In our analysis, the nodal non-Fermi liquid (NFL) behavior is also
evidenced by an anomalously high scattering rate $\Gamma_{\ve{k}_F}=
Z_{\ve{k}_F}|{\rm Im}\Sigma_0^{ph}|$ ($Z_{\ve{k}_F}$ is the calculated quasiparticle residue).
In Fermi-liquid theory of the free electron gas, one always finds $|{\rm
Im}{\Sigma}_0^{ph}| \ll E_2(\ve{k})$ sufficiently close to the Fermi
level which signifies the existence of coherent quasiparticles with a long life
time $\tau=1/\Gamma_{\ve{k}}$. At low $T$ and very close to the Fermi level,
the interactions of the free electrons to an Einstein phonon lead to
an exponential decrease of ${\rm Im}\Sigma_0^{ph} \sim \exp(-\beta\omega_0)$
as $T\rightarrow 0$.
This means that for small excitation energies $\varepsilon$ close to the Fermi level,
there always exists a low-temperature range for which $|{\rm
Im}{\Sigma}_0^{ph}| \ll \varepsilon$, which is consistent with the concept of
Landau quasiparticles~\cite{abrikosov}.
In contrast to this,
in the ordered system the nodal electron-phonon scattering near the Fermi level
produces a substantial
${\rm Im}\Sigma_0^{ph} \sim g^2 \{ \log|\Omega_2^S/t|+\log|\Omega_3^S/t| +
\eta_N(t,t',V)\}\exp(-\beta \omega_0)$ where the function $\eta_N(t,t',V)$ results
from a quadratic expansion of $\varepsilon_2(\ve{k})$ in the vicinity of the
nodal van-Hove singularities.
Due to the large dominant contributions of these van-Hove singularities,
in the temperature range $kT/t \le 0.03$ which is of relevance for the cuprates,
we obtain $|{\rm Im}{\Sigma}_0^{ph}| > \varepsilon$ where the excitation energies
$\varepsilon \le kT$ are located close to the Fermi level. The comparison of
the corresponding ${\rm Im}{\Sigma}_0^{ph}$ in the free and in the ordered
electron gas is presented in Fig.~\ref{fig4}. As a consequence, the strong increase
of ${\rm Im}{\Sigma}_0^{ph}$ and of the electron scattering rate $\Gamma_{\ve{k}_F}$
leads to a violation of the condition for well-defined quasiparticles in the electron-ordered
system.
We emphasize that the ultimate reason for such a
high electron-phonon scattering rate is the anomalously high density of
the nodal electronic states related to the flat electron dispersion.
The incoherent NFL nodal state which develops for $E_p>E_p^*$
is in striking contrast to the Fermi-liquid state with its
distinctive hole-type quasiparticle excitations for $E_p<E_p^*$ (dispersion
similar to $\varepsilon_2(\ve{k})$ in Fig.~\ref{fig2}(a)).
The existence of the singular multipole structure 
of the one-electron Green functions and of the anomalously high scattering
rates has been confirmed by more elaborate calculations which include
higher-order vertex corrections into the electron-phonon contribution to the electronic
self-energy. As follows from these calculations, the inclusion of
vertex corrections leads to slight shifts of the low-energy van-Hove singularities in ${\rm Im}\Sigma_0^{ph}$,
without qualitative changes in the nature of the NFL state.
In fact, evidence for a NFL behavior along the nodal region, which
is even unaffected by the onset of superconductivity, has been found for optimally doped BSCCO
\cite{valla}, in agreement with our findings.

\begin{figure}[t]
\epsfxsize=8.8cm \centerline{\epsffile{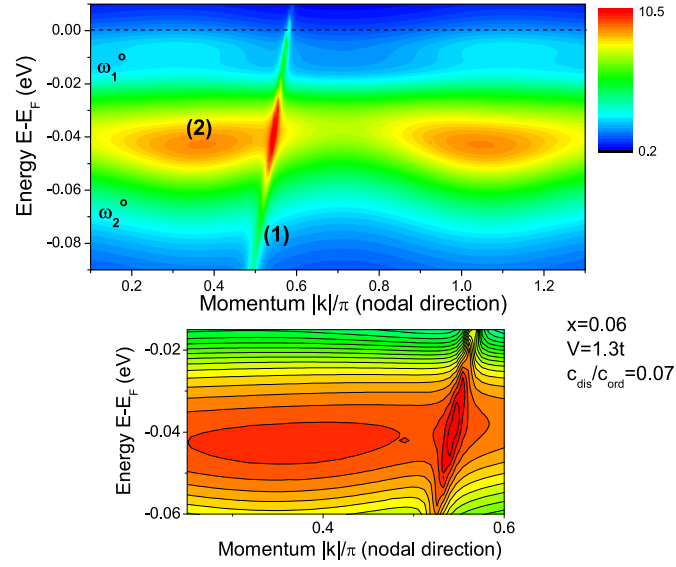}} \caption{Spectral intensity map
$I(\ve{k},\omega)$ for an inhomogeneous electron system with $x=0.06$ along the
nodal direction. The bottom panel displays the detailed structure of the
break-up feature at $\omega_2^o\approx -50$~meV Here the spectral intensity is
renormalized by the weight coefficient of the ordered state $c_{\rm ord}$;
$V/t=1.3$, $kT/t=0.03$, $\omega_0=0.05t$ and $E_p=1.2t$. The position of the Fermi level
is indicated by dashed line.} \label{fig5}
\end{figure}

\section{Electronic phase separation state}

As a state of electronic phase separation would imply a superposition of the ordered
and disordered electronic states on the analyzer,
the resulting ARPES intensity contains a combination of the disordered
($A^{d}(\ve{k},\omega)$) and the ordered ($A^{o}(\ve{k},\omega)$) spectral functions:
$I(\ve{k},\omega)=(c_{\rm ord}
A^{o}(\ve{k},\omega)+c_{\rm dis} A^{d}(\ve{k},\omega))f(\omega)$. Here
the coefficients $c_{\rm ord}$ and $c_{\rm dis}$
refer to the ordered and disordered surface fractions. We assume that
the ratio $c_{\rm dis}/c_{\rm ord}$ depends
linearly on doping so that $c_{\rm dis}/c_{\rm ord}\sim x$. The consequential
dominance of the charge disordered contribution for larger $x$ is consistent
with the expansion of metallic disordered regions at higher doping levels as
observed in Ref.~\cite{kohsaka}.

\begin{figure}[ht]
\epsfxsize=8.5cm \centerline{\epsffile{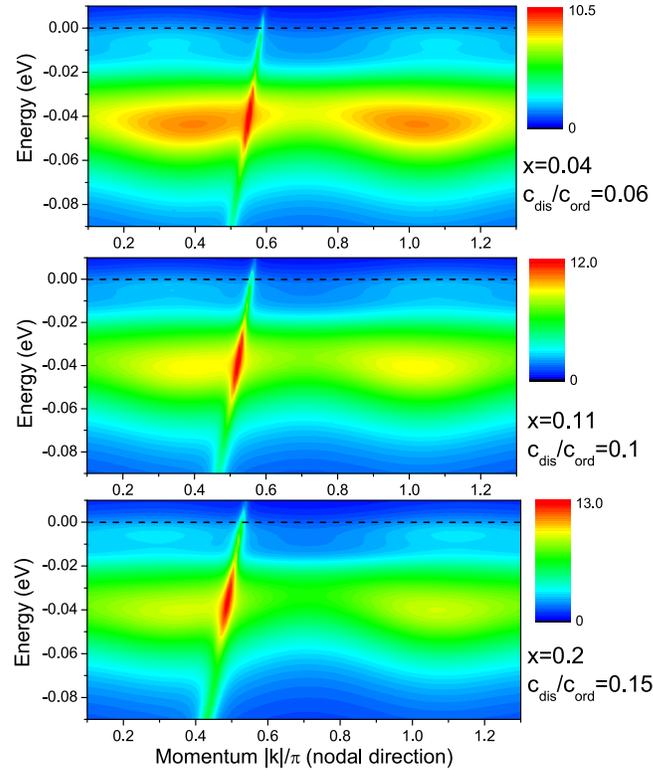}} \caption{Evolution of the spectral
intensity map $I(\ve{k},\omega)$ at different doping levels. The position of the Fermi level
is indicated by dashed line.
} \label{fig6}
\end{figure}

Fig.~\ref{fig5} shows the spectral intensity map in the nodal direction
calculated for $x=0.06$. One can clearly distinguish two structures on this
map. The needle shaped high intensity structure~(1) is related to the free
electronic band $\varepsilon_0$. Moreover, a symmetric broad feature (2)
appears, originating from the ordered state. The structure~(2) is located on
the plot in the range between $\omega^o_2 \approx -60$~meV and $\omega^o_1
\approx -10$~meV. The high spectral intensity of the structure (2) appears due
to the singular character of $E_2(\ve{k})$ in the vicinity of the point
${\rm S_1}$ of the nodal $n_1$-region in Fig.~\ref{fig2}(c). Near the Fermi level, the
NFL-region $n_1$ forms due to an additional van-Hove singularity at
$\omega=\omega_3$ (see Fig.~\ref{fig2}(b)) which results in multiple poles of
the electronic propagator $g_2$ and in high spectral intensity in the region
between $\omega_2^o$ and $\omega_1^o$. In the intensity $I(\ve{k},\omega)$, the
ordering-induced feature (2) is superimposed with the free band~(1) which
produces
a break-up of the intensity at $\omega\approx -50$~meV, shown in more
detail in the bottom panels of Fig.~\ref{fig5}. This break-up of the spectral
weight into a quasiparticle peak along the needle (1) and a broad high
intensity structure (2) is typical for the nodal ARPES measurements observed in
a wide variety of cuprate compounds. As the break-up is produced by the
charge-order gap, it should be considered as {\it a direct manifestation of the
local electron order in the cuprates}.

We note that the broad low-energy feature
(marked as part (2) in Fig.~\ref{fig5})
extends in the $\ve{k}$-direction which is in contrast to the experimental
observations. The reason for such a wide spread of this charge-order-induced
feature (2) is a simplified approximation for the characteristic phonon frequency
$\omega_0=const$. 
Within an Einstein approximation
the electron-phonon contribution to the electronic self-energy $\Sigma_{\pm}^{ph}(\omega)$
does not depend on the momentum vector $\ve{k}$, which in turn leads to a wide spread
of the self-energy part $\Sigma_2$ and of the corresponding high spectral intensity
region in the $\ve{k}$-space. To obtain 
better agreement with 
experiment, where the high-intensity structure is localized in 
$\ve{k}$-space, one needs to
consider a realistic description for the phonon dispersion which is beyond 
the scope of our current studies.

Due to the doping dependence of $\Sigma_2(\ve{k},\omega)$, the high intensity features
also depend strongly on doping which is demonstrated in Fig.~\ref{fig6}. The
high-intensity structure, at about $-50$~meV for $x=0.04$, becomes smoother
with increasing $x$ up to $x=0.11$ and in the overdoped regime ($x=0.2$
in Fig.~\ref{fig6}). Moreover, as the local order is destroyed with
increasing $T$,  the contribution of ordered domains to
$I(\ve{k},\omega)$ will decrease which can explain the fact of vanishing kinks
for higher $T$ discussed in
Refs.~\cite{kordyuk} and~\cite{johnson}.

The interpretation of ARPES intensities in terms of electronic inhomogeneities is a
possible scenario not only for cuprates, but can be applied also for manganite
compounds. In manganites,
a significant electron-phonon coupling and
an electronic phase separation of ferromagnetic
metallic and charge ordered states is strongly supported by numerous
experimental and theoretical studies \cite{bishop,moreo,dagotto}. Consequently, the
nodal ARPES features reported for La$_{1.2}$Sr$_{1.8}$Mn$_2$O$_7$ \cite{mannella}
can also be explained by the proposed approach for inhomogeneous states.

In conclusion, we show that coupling to phonon modes leads to different spectral
properties in ordered and
disordered electronic states. It appears to be the key mechanism responsible
for the main features detected in ARPES experiments of high-$T_c$
cuprates. The broad character of the ARPES features can be related to the
incoherent nature of the nodal non-Fermi-liquid state which
forms essentially due to the electron-phonon coupling. In this context, the
implications of electron-phonon coupling in charge ordered systems demonstrate
in a novel way how collective modes can qualitatively change the fundamental
properties of the electron liquid.

This work was supported by Deutsche Forschungsgemeinschaft through SFB~484.

\end{document}